\documentclass{aa}
\include{aas_macros}
\usepackage{color}
\usepackage{natbib}
\usepackage{graphicx}
\usepackage{txfonts}
\newcounter{IonCS}
\renewcommand{\ion}[2]{\setcounter{IonCS}{#2}#1\,{\small{\Roman{IonCS}}}}

\newcounter{RomC}

\newcommand{\be}{\begin{equation}}
\newcommand{\ee}{\end{equation}}
\newcommand{\ba}{\begin{eqnarray}}
\newcommand{\ea}{\end{eqnarray}}

\begin{document}
   \title{The multi-component field topology of sunspot penumbrae}
   \subtitle{A diagnostic tool for spectropolarimetric measurements}
   \author{D.A.N. M\"uller
          \inst{1}
          \and
          R. Schlichenmaier
          \inst{2}
          \and
          G. Fritz
          \inst{2,3}
          and
          C. Beck
          \inst{2}
          }

   \offprints{D.A.N. M\"uller}

   \institute{
     European Space Agency, Research and Scientific Support Department,\\ c/o NASA Goddard Space Flight Center, Mail Code 612.5, Greenbelt, MD 20771, USA\\
              \email{dmueller@esa.nascom.nasa.gov}
         \and
         Kiepenheuer-Institut f\"ur Sonnenphysik, Sch\"oneckstr. 6, 79104 Freiburg, Germany\\
\email{schliche@kis.uni-freiburg.de,cbeck@kis.uni-freiburg.de}
\and
Arnold Sommerfeld Center for Theoretical Physics, Theresienstr. 37, 80333 M\"unchen, Germany\\
\email{derfritz@gmx.net}
             }

   \date{Draft version}

\abstract
{Sunspot penumbrae harbor highly structured magnetic fields and flows. The moving flux tube model offers an explanation for several observed phenomena, e.g.\ the Evershed effect and bright penumbral grains. A wealth of information can be extracted from spectropolarimetric observations. In order to deduce the structure of the magnetic field in sunspot penumbrae, detailed forward modeling is necessary. On the one hand, it gives insight into the sensitivity of various spectral lines to different physical scenarios. On the other hand, it is a very useful tool to guide inversion techniques. In this work, we present a generalized 3D geometrical model that embeds an arbitrarily shaped flux tube in a stratified magnetized atmosphere. The new semi-analytical geometric model serves as a frontend for a polarized radiative transfer code. The advantage of this model is that it preserves the discontinuities of the physical parameters across the flux tube boundaries. This is important for the detailed shape of the emerging Stokes Profiles and the resulting net circular polarization (NCP). (a) The inclination of downflows in the outer penumbra must be shallower than approximately $15^\circ$. 
(b) Observing the limb-side NCP of sunspots in the \ion{Fe}{1}~1564.8\,nm line offers a promising way to identify a reduced magnetic field strength in flow channels. 
(c) The choice of the background atmosphere can significantly influence the shape of the Stokes profiles, but does not change the global characteristics of the resulting NCP curves for the tested atmospheric models.
}
 

   \maketitle
%

\section{Introduction}
Spectral  line profiles in the penumbra are characterized by line shifts and line asymmetries of their intensity (Stokes I) as well as in polarized light (Stokes Q, U, and V). A suitable measure of the asymmetry of Stokes-$V$ profiles is the area asymmetry or \emph{net circular polarization}, NCP, of a spectral line, which we define as
\be
{\rm NCP} \equiv \int\limits_{\delta \lambda} V(\lambda)\, {\rm d} \lambda \, , 
\ee
where the interval of integration, $\delta \lambda$, encompasses the whole line profile. This quantity has the advantage that it is independent of the spatial resolution of the observations since a convolution of a line profile with a Gaussian preserves its area.
 The first measurements of the net circular polarization in sunspots were reported by \cite{Illing+al1975}.
\cite{Auer+Heasley1978} showed that these observations could be explained by assuming macroscopic velocity fields. Furthermore, they proved that velocity gradients are a necessary \emph{and} sufficient condition to produce a NCP.

The advent of modern polarimeters like the Advanced Stokes Polarimeter \citep[ASP, ][]{Skumanich+al1997ApJS} at the NSO Vacuum Tower Telescope in Sunspot, New Mexico, and the Tenerife Infrared Polarimeter \citep[TIP, ][]{MartinezPillet+al1999ASP} and the Polarimetric Littrow  Spectrometer \citep[POLIS, ][]{Beck2005AA_POLIS} at the Vacuum Tower Telescope (VTT) on Tenerife has brought new challenges to observers as well as theoreticians. It is now possible to routinely record maps of the full Stokes vector for a whole sunspot. This facilitates the study of spectral signatures of magnetic flux tubes in the penumbra by analysis of maps of the NCP.

So far, the spectral synthesis has been limited to atmospheric models based on snapshots of the moving tube model simulations by \cite{Schlichenmaier1998} and to configurations in a two-dimensional plane (``slab models''). Such slab models have been used earlier by \citet{solanki+montavon1993}, \citet{Sanchez+al1996} and \citet{MartinezPillet2000}, and led to the picture of the ``uncombed'' penumbra: Approximately horizontal magnetic flux tubes are embedded in a more inclined background field that is at rest. The sharp gradient or even discontinuity of the velocity of the plasma between the inside of the tube and the static surroundings result in the observed NCP. While two-dimensional models reproduced the observed NCP and its center-to-limb variation, they are limited to the few geometric cases where a flux tube is exactly aligned with the line-of-sight  (LOS). Thus, they do not give any information about the NCP at different positions within the penumbra.

\begin{figure}[ht]
\begin{center}
\resizebox{0.8\hsize}{!}{
\includegraphics{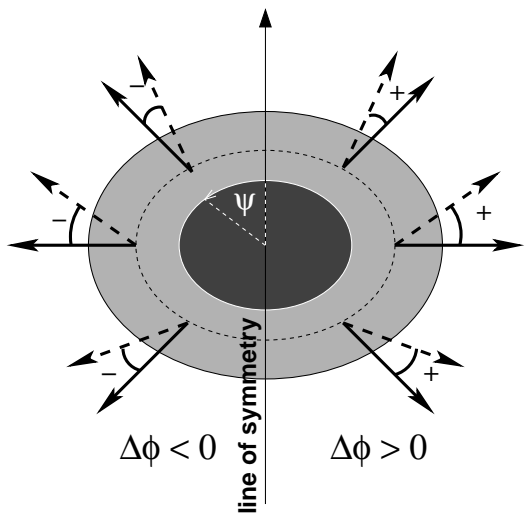}
}
\end{center}
\caption{\label{gfx_los}Pictorial representation of the difference in azimuth, $\Delta \phi(\psi)$, between the magnetic field vector inside a flux tube (solid arrows) and that of the background field (dashed arrows). The tubes are nearly horizontal, while the field vector of the background has a steeper inclination, but is also in the plane spanned by the spot axis and the flux tube axis.
}
\end{figure}

We showed \citep[][hereafter M02]{ncpletter2002, Mueller+al2002} that a wealth of additional information is contained in NCP maps, which can be retrieved with the help of a three-dimensional model.
Discontinuities in the azimuth of the magnetic field (as measured in the observer's coordinate system) lead to a characteristic pattern of the NCP within the penumbra. Specifically, it was shown that the apparent symmetry within the spot is broken due to the following fact: The discontinuities in the magnetic field azimuth are different for two locations that are symmetrically located in the spot with respect to the ``line-of-symmetry'', i.e.\ the line connecting disk-center and sunspot center (Fig.~\ref{gfx_los}). These azimuth discontinuities have been confirmed observationally by \citet{BellotRubio+al2004AA}.

The impact of this geometric effect on the NCP strongly varies between spectral lines at different wavelengths. 
A comparison of the NCP of different spectral lines can thus be used to diagnose the magnetic field inclination and orientation of flow channels in the penumbra. As an example, we consider the two iron lines \ion{Fe}{1}~1564.8\,nm and \ion{Fe}{1}~630.25\,nm. Both lines are Zeeman triplets and have similar Land\'e factors of g=3 and g=2.5, respectively.
It turns out that the infrared line (\ion{Fe}{1}~1564.8\,nm) is far more susceptible to the symmetry breaking than the visible line (for details, cf.\ M02).

\section{The VTUBE model}\label{sec_model}

In order to gain a better understanding of the different factors that determine the NCP and its spatial variation within the penumbra, we have constructed a 3D geometric model (VTUBE) of a magnetic flux tube embedded in a background atmosphere. This model serves as the frontend for a radiative transfer code \citep[DIAMAG,][]{ugd1994}. Combining the two, we can generate synthetic NCP maps for any desired axisymmetric magnetic field configuration and arbitrary flux tube properties.
The model has been built to offer a high degree of versatility, e.g.\ the option to calculate several parallel rays along the line-of-sight that intersect the tube at different locations. One can then average over these rays to model observations of flux tubes at different spatial resolutions and for different magnetic filling factors of the atmosphere. Furthermore, we can also take into account radial variations of the physical properties of the flux tube. Doing so, one can e.g.\ model the interface between the flux tube and its surroundings.

In the simplest case, the VTUBE model reduces to a general model for a two-component atmosphere, i.e.\ an atmosphere with a static background component and a flow component. Thus, it can also be used for flux tube models other than the moving tube model considered here.

\begin{figure}[ht]
\resizebox{\hsize}{!}{
\includegraphics{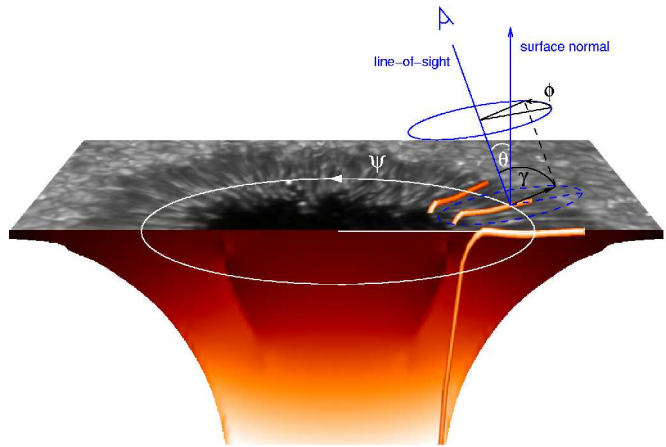}}
\caption{\label{gfx_azimut_sketch} Magnetic flux tubes in the penumbra. For a given heliocentric angle $\theta$, the azimuth $\phi$ and the inclination $\gamma$ of the magnetic field vary with the location of the flux tube in the sunspot, characterized
by the spot angle $\psi$.}
\end{figure}

Figure \ref{gfx_azimut_sketch} shows the relevant angles that appear in the model. For a given heliocentric angle $\theta$, the azimuth $\phi$ and the inclination $\gamma$ of the magnetic field vary with the location of the flux tube in the sunspot, characterized by the spot angle $\psi$.
For the calculations presented in this paper, no averaging over parallel rays was performed, and the physical quantities were kept constant along the flux tube radius (thin flux tube approximation).

\subsection{The background atmosphere}

Apart from the details of the magnetic flux tube, the VTUBE model allows the user to specify the background atmosphere. We assume that the thermodynamic part of the atmospheric model only depends on the geometrical height, $z$. Thus, temperature, $T$, pressure, $p$, and density, $\rho$, are prescribed in tabulated form, with data obtained either from semi-empirical or theoretical models. The magnetic field structure is described independently from the thermal structure. This is done because we are interested in the effect of varying physical parameters of the flux tube, and there is so far no self-consistent 3D MHD model of flux tubes in the penumbra available. For atmospheric models for which the magnetic vector field has not been determined, one has to take care that the parametrization of the magnetic field is consistent with the thermodynamic structure.

We use the same parametrization as in M02 to model an axisymmetric magnetic field whose strength decreases radially as
$
B (r) = B_0 / (1+(r/r_*)^2) 
$
and whose inclination decreases linearly as
$
\gamma^\prime (r) = \pi r /(2 r_{\gamma^\prime}) \,.
$
In these expressions, $r$ denotes the radial distance from the spot center, $r_*$ is the radial distance at which the magnetic field has decreased by 50\%, and $\gamma^\prime$ indicates the radial distance at which the field becomes horizontal. This represents an extension to the work of \cite{BS1969} which takes into account more recent observations \citep[e.g.][]{Title1993}, according to which the magnetic field of a sunspot is not horizontal at the spot boundary ($r = r_*$, where $r_*$ is the radius at which the field strength has dropped to half its maximal value), but shows a small inclination with respect to the surface. Adopting values of $B_0 = 2700$\,G, $r_* = 16\,800$\,km, and $r_{\gamma'} = 17\,400$\,km yields an inclination of  $3.3^{\circ}$ at $r=r_*$, and a vertical gradient of the magnetic field in the spot center of $\partial B_z/\partial z\approx - 0.25$\,G/km, which agrees fairly well with the value of $\partial B_z/\partial z\approx - 0.3$\,G/km as determined from the model of \citep[][hereafter JS94]{Jahn+Schmidt1994} for $r=0$.

In this work, we experiment with three different background atmospheres. The first one is the penumbral model of JS94 which was used as a background atmosphere for the MHD simulations of \cite{Schlichenmaier1998} and for the NCP calculations of M02.
 The second one is the penumbral model of \cite{BellotRubio+al2006AA} which is based on high-resolution intensity profiles of the non-magnetic \ion{Fe}{1} 557.6\,nm line, obtained with the TESOS spectrometer at the Vacuum Tower Telescope on Tenerife. The third one is the quiet sun model of \cite{Holweger+Mueller1974} modified by \cite{SchleicherPhD} to account for the chromospheric temperature rise. This model was chosen to study the effect of a slightly higher temperature in the line-forming region of the atmosphere.

\begin{figure*}[ht]
\begin{center}
\includegraphics[width=0.7\hsize]{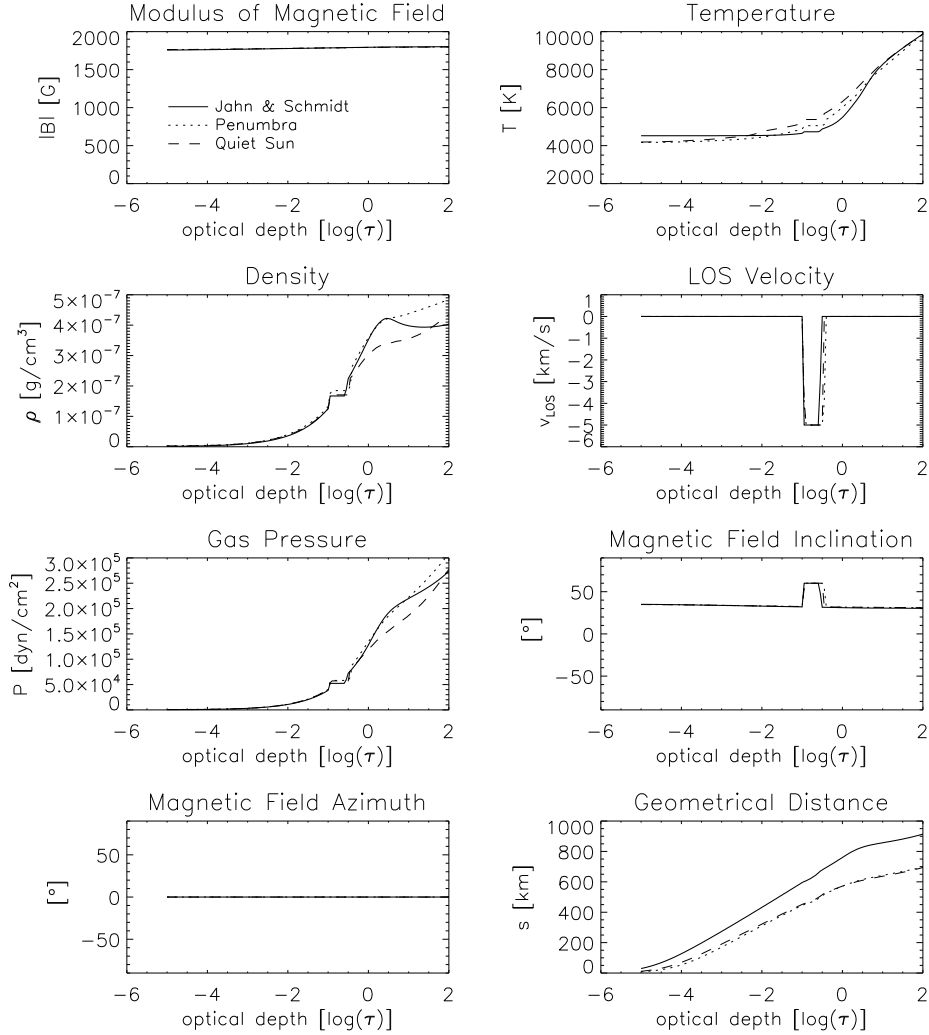}
\end{center}
\caption{\label{gfx_atmpar}
Physical parameters along the line-of-sight for three different atmospheric models. The geometrical location of the flux tubes has been adjusted so that the top of the flux tube is located at $\log(\tau) = -1$ for a vertical line-of-sight ($\theta = 0^\circ$) for all three models. The example shows a magnetic flux tube with a flux of ${\cal  F}= 5\cdot 10^{16}$\,Mx in the outer penumbra ($r = 12\,000$\,km), observed at a heliocentric angle of $\theta = 30^\circ$ and a spot angle of $\psi = 0^\circ$. The flow speed inside the tube is $v_0 = 10$\,km/s which results in a line-of-sight velocity of $v_{LOS} = v_0 \cdot \sin(\theta) = 5$\,km/s. Since the flux tube is aligned with the line-of-sight, the azimuth angle is zero.
}
\end{figure*}

Figure~\ref{gfx_atmpar} shows the physical parameters along the line-of-sight for the three different atmospheric models with a horizontal magnetic flux tube.
The upper right panel shows the temperature of these three atmospheric models as a function of optical depth. Between $\log(\tau)=-2$ and $\log(\tau)=0.5$, the quiet sun model is up to 570\,K hotter than the model of \cite{BellotRubio+al2006AA} and up to 860\,K hotter than the atmosphere of JS94.

Since spectra yield information about the atmospheric parameters as a function of optical rather than geometrical depth and since the NCP is very sensitive to the exact location of the flow channel along the line-of-sight, the location of the flux tube has been adjusted to allow a comparison between the different models. The top of the tube was placed at a relative height of $z_t=-50$\,km for the JS94 model, $z_t=+82$\,km for the penumbral model of \cite{BellotRubio+al2006AA} and $z_t=+78$\,km for the quiet sun model of \cite{SchleicherPhD}, so that the top of the flux tube is located at an optical depth of $\log(\tau) = -1$ for a vertical line-of-sight ($\theta = 0^\circ$). The example shows a magnetic flux tube with a flux of ${\cal  F}= 5\cdot 10^{16}$\,Mx in the outer penumbra ($r = 12\,000$\,km), observed at a heliocentric angle of $\theta = 30^\circ$ and a spot angle of $\psi = 0^\circ$ (i.e.\ center-side penumbra). The flow speed inside the tube is $v_0 = 10$\,km/s (this speed is used for all calculations in this paper unless otherwise noted) which results in a line-of-sight velocity of $v_{LOS} = v_0 \cdot \sin(\theta) = 5$\,km/s. The minus sign indicates that the flow is directed towards the observer.  Since the flux tube is aligned with the line-of-sight, the azimuth angle is zero.

\subsection{Building a realization of the model}

Once the background atmosphere has been defined, the following steps are performed to assemble a realization of the penumbral flux tube model:
\begin{enumerate}
\item The flux tube axis is defined by a polygon or spline function.
\item The physical properties of the flux tube are defined and then interpolated along the flux tube axis.
\item The viewing angle is defined by specifying the heliocentric angle, $\theta$, and the location of the line-of-sight.
\item The spectral line profiles are calculated.
\end{enumerate}

\noindent These steps are described in detail in the appendix.

\section{From Stokes-V profiles to NCP curves}

\begin{figure*}[ht]
\begin{center}
\resizebox{0.6\hsize}{!}{
\includegraphics{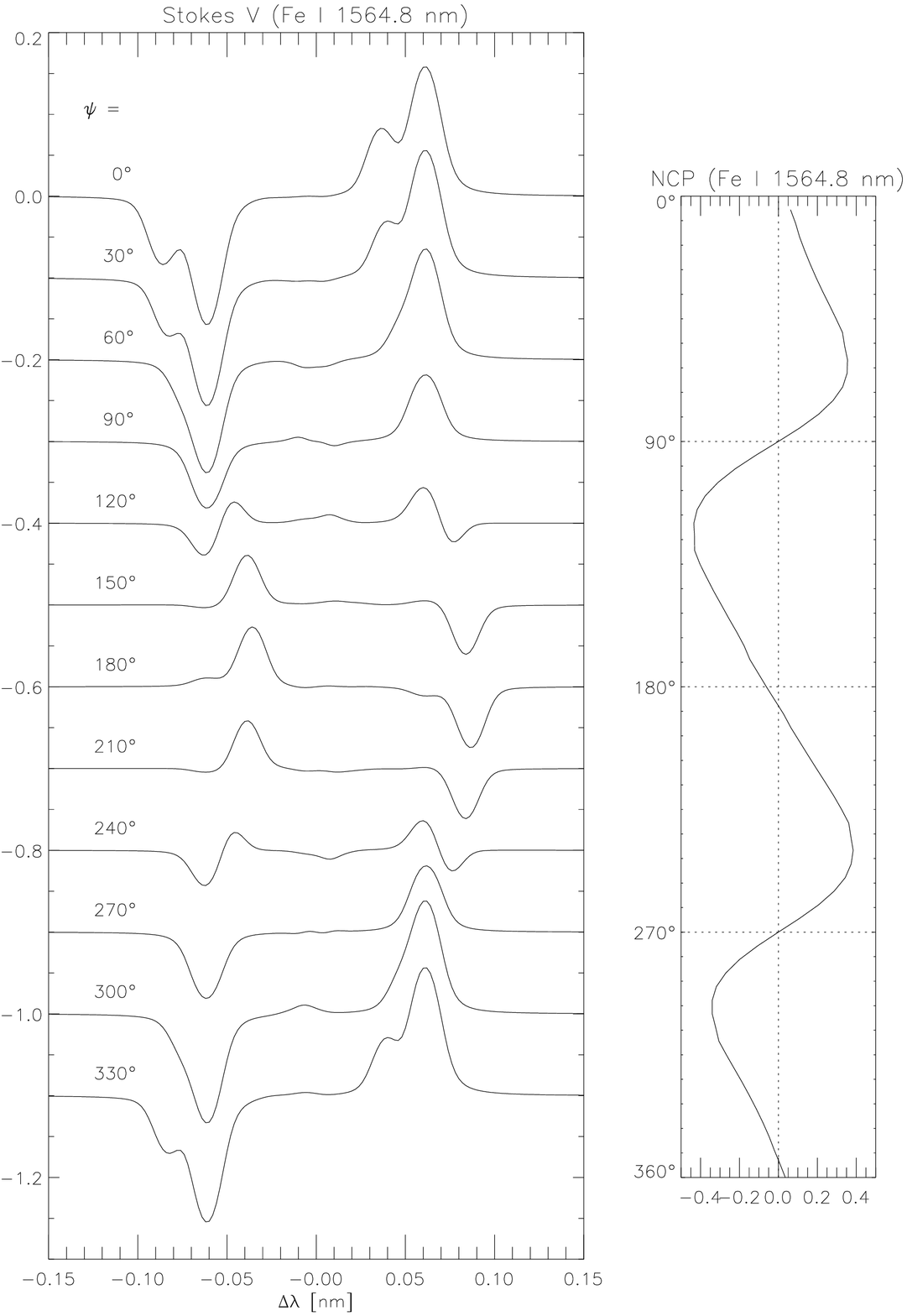}
}
\end{center}
\caption{\label{gfx_vstack}Stokes-V profiles and NCP for a horizontal magnetic flux tube with a flux of ${\cal  F}= 5\cdot 10^{16}$\,Mx, embedded in the penumbral atmospheric model of \cite{BellotRubio+al2006AA}, observed at a heliocentric angle of $\theta = 30^\circ$. The stack plot on the left side shows Stokes-V profiles for different positions along a radial cut at $r=12\,000$\,km through the outer penumbra, specified by the spot angle $\psi$. The right plot shows the corresponding NCP, i.e.\ the integral over the Stokes-V profiles. As expected, it is seen that the NCP vanishes for $\psi=90^\circ$ and $\psi=270^\circ$, where the line-of-sight is perpendicular to the flow.}
\end{figure*}

To illustrate how the shape of Stokes-V profiles is connected to a NCP curve, Fig.~\ref{gfx_vstack} shows Stokes-V profiles and the corresponding NCP for a straight horizontal magnetic flux tube. The tube contains a magnetic flux of ${\cal  F}= 5\cdot 10^{16}$\,Mx, is embedded into the penumbral atmospheric model of \cite{BellotRubio+al2006AA} and is observed at a heliocentric angle of $\theta = 30^\circ$. The stack plot on the left side shows Stokes-V profiles for different positions along a radial cut at $r=12\,000$\,km through the outer penumbra, specified by the spot angle, $\psi$. The spot angle runs counter-clockwise from the line connecting the center of the spot and the center of the solar disk (see Fig.~\ref{gfx_azimut_sketch}). For example, $\psi = 0^\circ$ corresponds to a position in the center-side penumbra where the Evershed flow is directed towards the observer. This can be seen by the blue-shifted component of the Stokes-V profile. The right plot shows the corresponding NCP, i.e.\ the integral over the Stokes-V profiles. As expected, it is seen that the NCP vanishes for $\psi=90^\circ$ and $\psi=270^\circ$, where the line-of-sight is perpendicular to the flow.

The Zeeman splitting of a spectral line is proportional to the square of its wavelength, $\lambda$, while the Doppler shift increases only linearly with $\lambda$. Due to this elementary difference, the effect of a flow channel in the line-of-sight of a magnetized atmosphere is most clearly seen for longer wavelengths, e.g.\ in the infrared. In the upper and lower profiles in Fig.~\ref{gfx_vstack} ($\psi \in [330^\circ,30^\circ]$), the flux tube component of the Stokes-V profile is clearly visible as a small hump (or ``line satellite'') in the wing of the background component. In this case, the observer is looking at the center-side penumbra. For $\psi = 90^\circ$ and $270^\circ$, the NCP vanishes since the line-of-sight is perpendicular to the flow. The azimuthal variation of the NCP is approximately antisymmetric with respect to the line connecting spot center and disk center. This behavior is characteristic of the infrared line \ion{Fe}{1} 1564.8\,nm and is known as the  \emph{$\Delta\phi$ effect} (see Sect.~\ref{sec_alpha} and M02 for details).

\section{Results}

\subsection{NCP maps}

\begin{figure*}[ht]
\begin{center}
\resizebox{\hsize}{!}{
\includegraphics{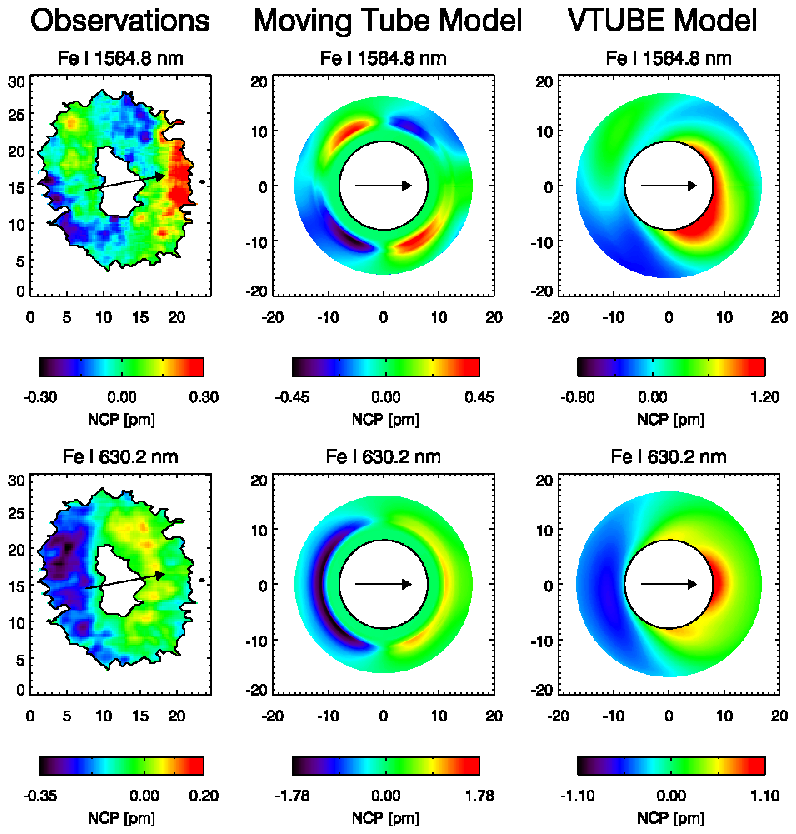}
}
\end{center}
\caption{\label{gfx_comp}Maps of net circular polarization (NCP) for an heliocentric angle of $\theta=30^\circ$. Upper row: infrared line, bottom row: visible line. Columns from left to right: observations (VTT, Tenerife), NCP maps synthesized from the moving tube model, and maps from the new and generalized VTUBE model. The arrows point towards disk center. The absolute magnitude of the NCP is higher for the synthetic maps than for the observations since a filling factor of unity is assumed. The color coding for all maps represents the NCP, measured in pm, and the spatial scales are  in Mm.}
\end{figure*}

\begin{figure}[t]
\includegraphics[width=\hsize]{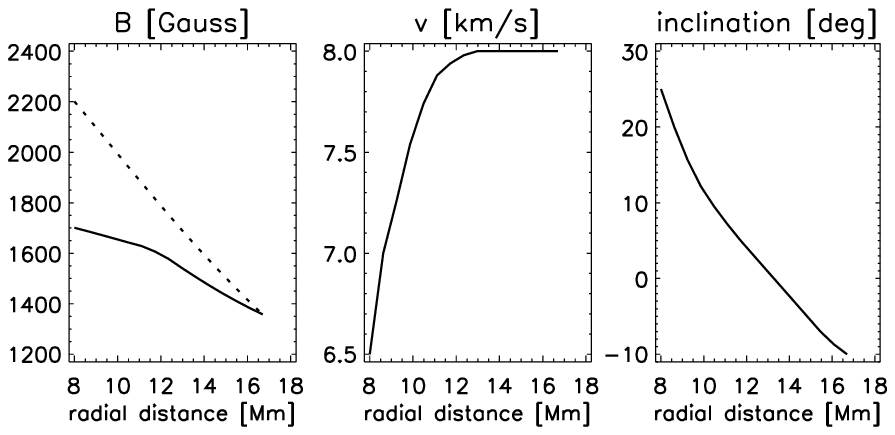}
\caption{\label{gfx_topology} Radial dependence of flow and magnetic field of the two components. These values are used to construct the NCP maps in the right panel of Fig.~\ref{gfx_comp}. The left panel shows the radial variation of the magnetic field strength of the strong background (dotted line) and weak tube (solid line) component. The middle and right panels display the flow speed and the inclination of the tube component relative to the horizontal.}
\end{figure}

Figure~\ref{gfx_comp} shows a comparison between NCP maps from observations, the moving tube model, and a map that is constructed with the new VTUBE model.
The upper row shows NCP maps of the \ion{Fe}{1} 1564.8\,nm infrared line, the bottom row shows NCP maps of the \ion{Fe}{1} 630.25\,nm visible line:
\begin{enumerate}
\item
The first set of maps is constructed from measurements that have been performed at the German Vacuum Tower Telescope (VTT), feeding two polarimeters simultaneously: the Tenerife Infrared Polarimeter \citep[TIP, ][]{MartinezPillet+al1999ASP}
for \ion{Fe}{1} 1564.8\,nm and the Polarimetric Littrow  Spectrometer \citep[POLIS, ][]{Beck2005AA_POLIS} for \ion{Fe}{1} 630.2\,nm.
The spot was at a heliocentric angle of $\theta=30^\circ$. More details on the observations are given in \cite{Beck2005PhD}.
\item
The second set of NCP maps is based on the moving tube model, as described in detail in M02.
\item
The third set of maps demonstrates the capability of the new VTUBE model. For the atmospheric parameters, we use results from two-component Stokes inversions. Analyses of such inversions have demonstrated that two interlaced components are capable of reproducing the spectropolarimetric data sets \citep[cf.~][]{Borrero+al2006AA,BellotRubio+al2004AA,Beck2005PhD}. The typical properties of these two components are summarized in Fig.~\ref{gfx_topology} and used as input for VTUBE: From the inner to the outer penumbra, the magnetic field strength of the tube component is about 500\,G smaller than that of the background component and gradually approaches it as the radial distance from spot center increases. The flow speed increases from 6.5 to 8 km/s in the inner penumbra and stays constant in the outer penumbra. The tube inclination gradually changes from being upwards ($\alpha=25^\circ$) to downwards  ($\alpha=-10^\circ$), with $\alpha$ being measured relative to the horizontal. Assuming that these parameters are axially symmetric, we employed the VTUBE model to calculate synthetic spectra for different viewing angles for the entire spot. For the background atmosphere, we used the penumbra model of \cite{BellotRubio+al2006AA}, and the tube was placed 50 km above optical depth unity with a magnetic flux of $5\times10^{16}$\,Mx. The resulting NCP maps are shown in the right column of Fig.~\ref{gfx_comp}.
\end{enumerate}

The observations (left column) indicate that the NCP distribution around the penumbra has two maxima and two minima in the infrared and is roughly antisymmetric with respect to the line-of-symmetry (line from spot center to disk center, indicated by a red arrow).
The moving tube model (second column from the left) is able to reproduce this fundamental difference. 
Since it is limited, however, to given snapshots of the MHD model, VTUBE serves to produce maps for other parameter sets. In the third column we use a set of parameters as inferred from inversions of spectropolarimetric data \citep{Beck2005PhD}.
The corresponding NCP maps reproduce the general symmetry properties, but have different radial dependences than the maps resulting from the moving tube model. When compared to observations it is seen that the radial dependence on the limb-side is well reproduced, but that the center-side penumbrae differ in terms of radial dependence while the azimuthal dependence is comparable.
In the case of the moving tube model, the sharp edges of the NCP contours in the inner penumbra are due to the emerging flux tube which enters the photosphere  and thus suddenly becomes visible in the two chosen spectral lines. For the VTUBE model calculations (right column), flux tube parameters inferred from observations were used which indicate more gentle inclination angles of the flux tube component. This explains the smoother contours for the latter model. 
The absolute magnitude of the NCP is higher for the synthetic maps than for the observations since a filling factor of unity is assumed, i.e.\ each line of sight is passing through the center of a magnetic flux tube. A recent analysis by \cite{Beck2005PhD}, based on Stokes inversions, indicates that the filling factor of the flow channel  component is on the order of 40\%. Since the NCP scales linearly with the filling factor, adopting this value significantly improves agreement between the magnitude of the observed NCP and the one calculated with the new VTUBE model. However, the work of \cite{Beck2005PhD} also indicates that the filling factor varies throughout the penumbra, so a detailed analysis is needed in order to obtain a more qualitative comparison.
In general, these results are very promising and have stimulated further modeling efforts. One open question is whether magnetic flux tubes dive back down into lower atmospheric layers in the outer penumbra and, if so, at which angle.

\subsection{\label{sec_alpha}The inclination of penumbral flux tubes}
\begin{figure*}[ht]
\begin{center}
\includegraphics[width=8cm]{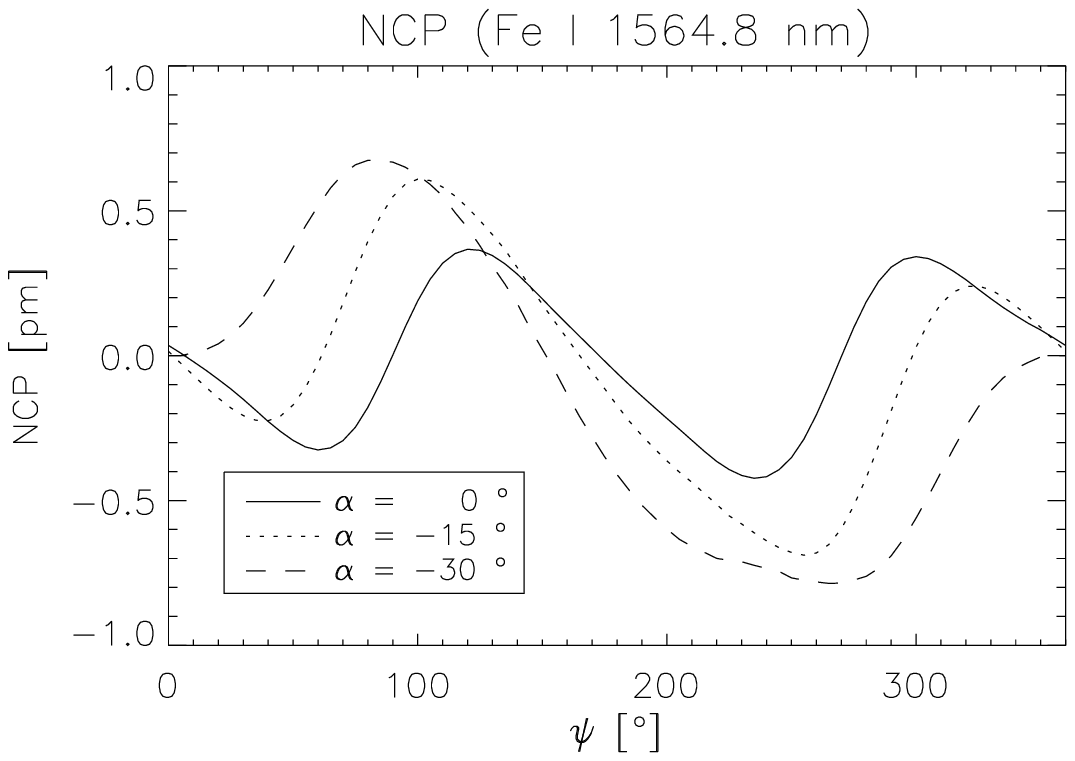}
\includegraphics[width=8cm]{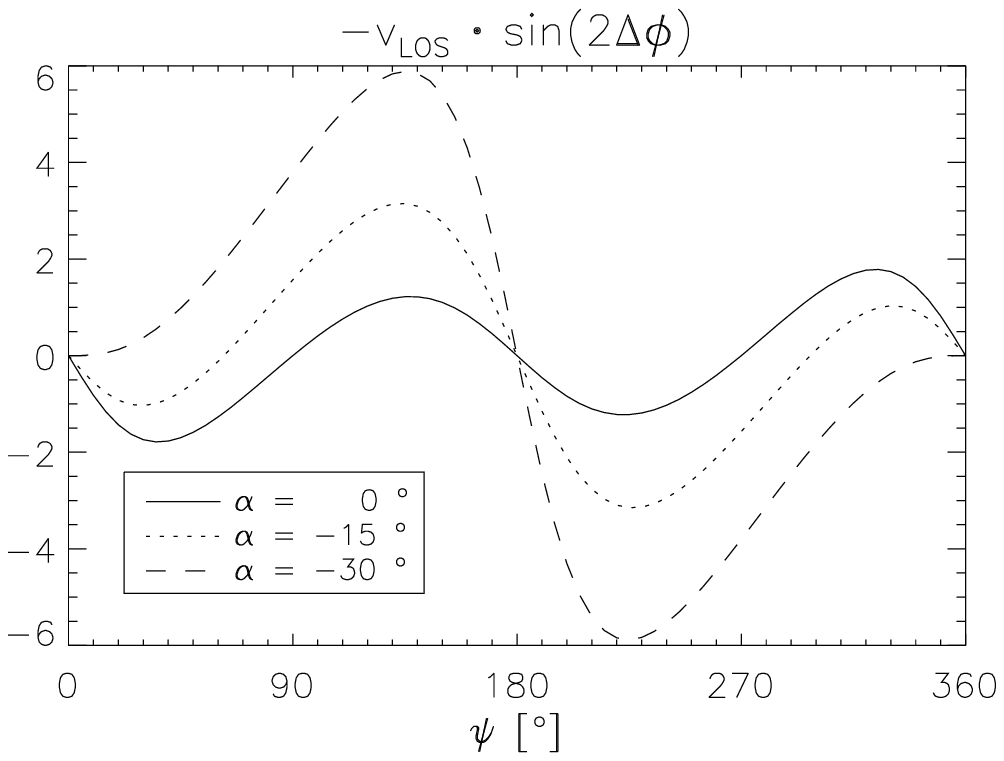}
\end{center}
\caption{\label{gfx_alpha}Variation of the NCP along a circular cut in the outer penumbra for the infrared line \ion{Fe}{1} 1564.8\,nm. 
For tube inclination angles of $\alpha \ge -15^\circ$, the curve has two pronounced maxima and minima, while there is only one of each for smaller angles, i.e.\ more vertical downflows.
The different shapes of the curves can be approximated by the term $-v_{LOS} \cdot \sin(2\Delta\phi)$ where $v_{LOS}$ is the line-of-sight velocity and $\Delta\phi$ is the jump in the azimuth of the magnetic field between the flux tube and the background atmosphere.}
\end{figure*}

Figure~\ref{gfx_alpha} shows the variation of the NCP along a circular cut in the outer penumbra for the infrared line \ion{Fe}{1} 1564.8\,nm. 
For tube inclination angles of $\alpha \ge -15^\circ$, the curve has two pronounced maxima and minima, while there is only one of each for smaller angles, i.e.\ more vertical downflows.

There are two points to be made here. The first one is that the different shapes of the curves can be explained by applying the simplified analytical model of \citet{Landolfi+landi1996}. In this model, a thin slab of moving plasma is located on top of a semi-infinite static atmosphere, and the two parts are permeated by different magnetic fields. In M02 we showed that their findings can also be applied to the slightly more complex situation of a magnetic flux tube embedded in a static background atmosphere. The NCP resulting from such an atmosphere with a flow channel can be caused by the gradient or jump in (a) the magnetic field strength ($\Delta B$ effect), (b) its inclination ($\Delta \gamma$ effect) and/or (c) the azimuth ($\Delta \phi$ effect). As described in detail in paper M02, we showed that the $\Delta \phi$ effect is the dominant one for the infrared line \ion{Fe}{1} 1564.8\,nm. In this case, the NCP is proportional to $-v_{LOS} \cdot \sin(2\Delta\phi)$ where $v_{LOS}$ is the line-of-sight velocity and $\Delta\phi$ is the jump in the azimuth of the magnetic field between the flux tube and the background atmosphere. This term is plotted in the right panel and qualitatively reproduces the symmetry properties of the NCP displayed in the left panel.

 The second point we would like to make is that the observed azimuthal variations of the NCP in the infrared line always show two maxima and two minima. Comparing this with models where the angle of the flux tube was varied, this finding rules out flux tubes that descend at an angle steeper than about $\alpha < -15^\circ$.

\subsection{Effect of lower magnetic field strength in the flux tube}

\begin{figure}[ht]
\resizebox{\hsize}{!}{
\includegraphics{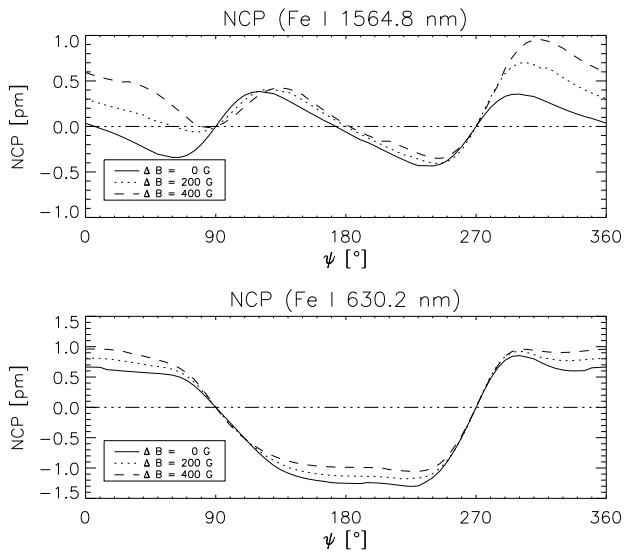}
}

\caption{\label{gfx_deltaB}
Introducing a jump  $\Delta B = B_{BG}-B_{tube}$ in the magnetic field strength between the flux tube and the background increases the NCP. For the infrared line (upper panel), the effect is largest for the center-side penumbra.}
\end{figure}

\begin{figure*}[ht]
\begin{center}
\resizebox{0.7\hsize}{!}{
\includegraphics{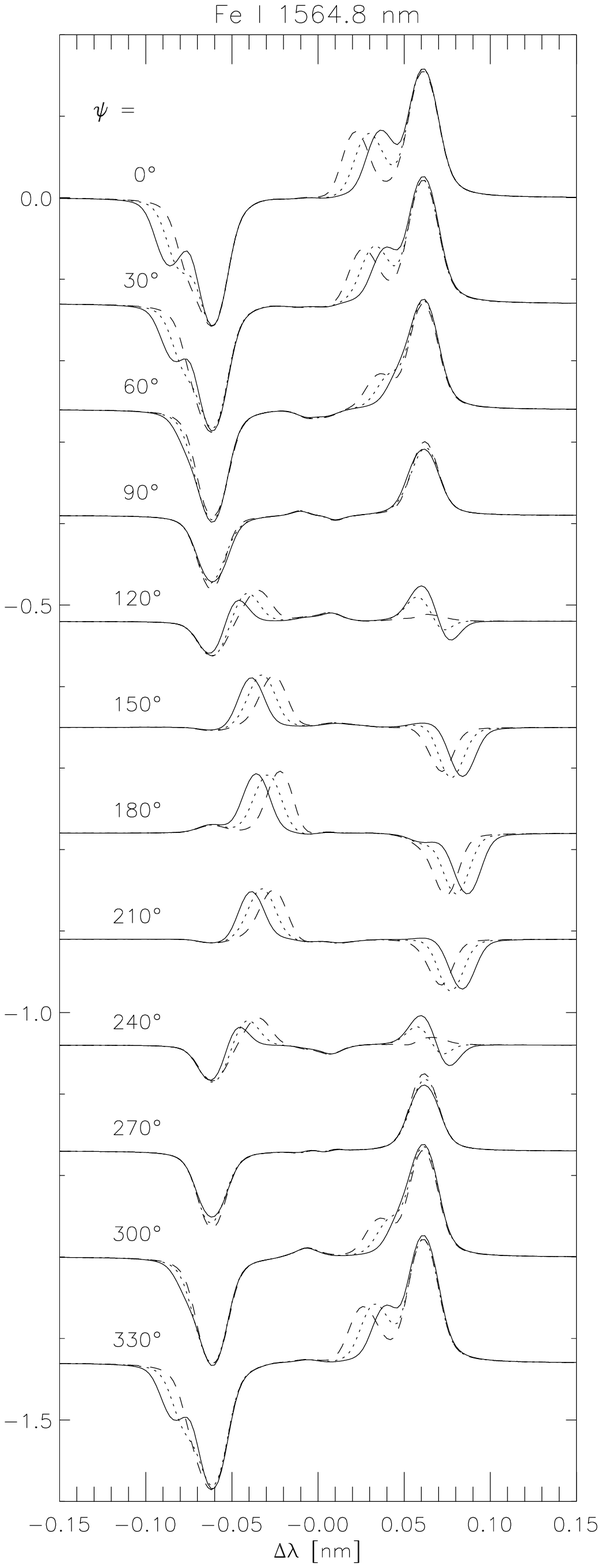}
\includegraphics{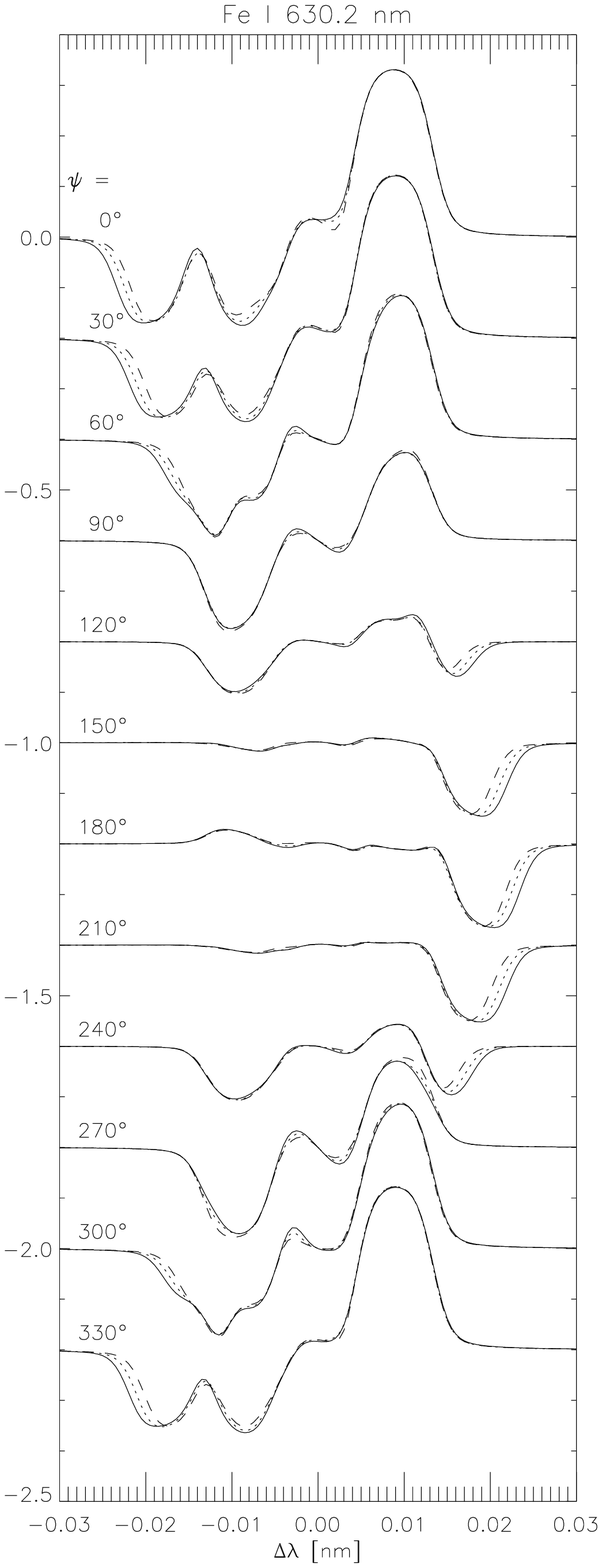}
}
\end{center}
\caption{\label{gfx_vstack_deltaB}
The effect of a jump  $\Delta B = B_{BG}-B_{tube}$ in the magnetic field strength between the flux tube and the background. The left panel shows the Stokes-V profiles of the infrared line, the right panel the Stokes-V profiles of the visible line. The solid line corresponds to $\Delta B = 0$\,G, the dotted line to  $\Delta B = 200$\,G and the dashed line to  $\Delta B = 400$\,G.
}
\end{figure*}

Recent observations by \cite{BellotRubio+al2004AA} and \cite{Borrero+al2004AA} suggest that the magnetic field strength in penumbral flux tubes is lower than in its surroundings.
One way of confirming this conjecture is to look for the signature of a discontinuity in the magnetic field strength in the NCP.
Figure~\ref{gfx_deltaB} shows the azimuthal variation of the NCP for both the infrared and the visible line for the atmospheric model of \cite{BellotRubio+al2006AA}. The effect of lowering the magnetic field strength inside the flux tube is the following: For the infrared line \ion{Fe}{1}~1564.8\,nm, the NCP strongly increases in the center-side penumbra ($\psi \in [270^\circ, 90^\circ]$). The dotted lines in the plots corresponds to a jump in magnetic field strength of $\Delta B = B_{BG}-B_{tube} = 200$\,G, the dashed lines corresponds to  $\Delta B = 400$\,G.
For the visible line \ion{Fe}{1}~630.25\,nm, the effect is much smaller and consists of a small increase of the NCP in the center-side penumbra and a small decrease of the NCP in the limb-side penumbra.
The explanation for this behavior is given in Fig.~\ref{gfx_vstack_deltaB}: The left panel shows the Stokes-V profiles of the infrared line, the right panel the Stokes-V profiles of the visible line (same line-styles as in Fig.~\ref{gfx_deltaB}). Let us take a look at the center-side profiles around $\psi=0^\circ$: It can be seen that upon reduction of the field strength inside the tube, the two small humps of the flux tube component of the profile move closer together due to the reduced Zeeman effect. For the blue wing of the Stokes-V profile, this means that the small hump merges with the main lobe, which reduces the integrated area over the blue wing. For the red wing, on the contrary, the small hump becomes more and more disconnected from the main lobe. This increases the integrated area under the red wing, with the result that the total integral over the Stokes-V profile increases. In the case of the visible line, reducing the magnetic field strength increases the NCP just a little on the center-side (because the flux tube component of the Stokes-V profile is less split due to the Zeeman effect). On the limb-side, the NCP is slightly reduced for the same reason (compare the profiles in the right panel of Fig.~\ref{gfx_vstack_deltaB}).

In conclusion,  observing the limb-side NCP of sunspots in the \ion{Fe}{1}~1564.8\,nm line offers a promising way to identify a reduced magnetic field strength in flow channels. If the NCP is significantly different from zero and does not show antisymmetry around the line connecting spot center and disk center, this is a strong hint for a lower magnetic field strength in the flux tubes. 

\subsection{Effect of different background atmospheres}

\begin{figure*}[ht]
\begin{center}
\resizebox{0.7\hsize}{!}{
\includegraphics{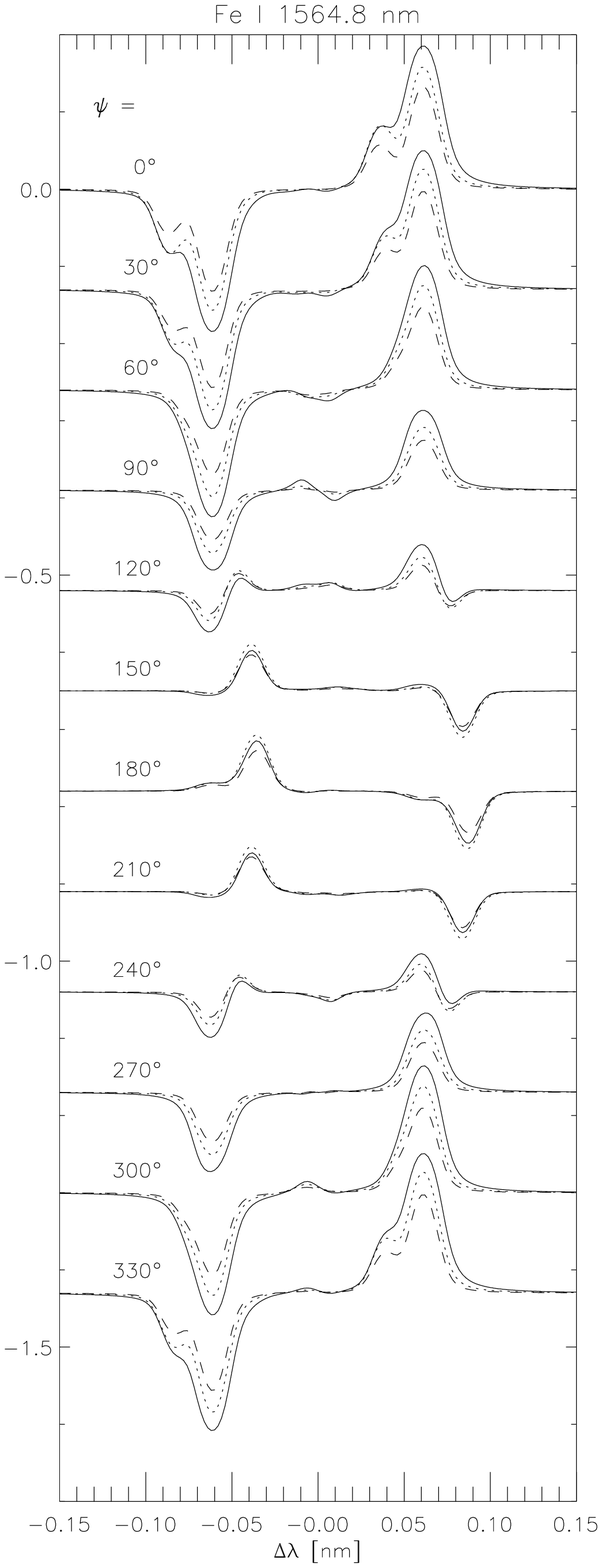}
\includegraphics{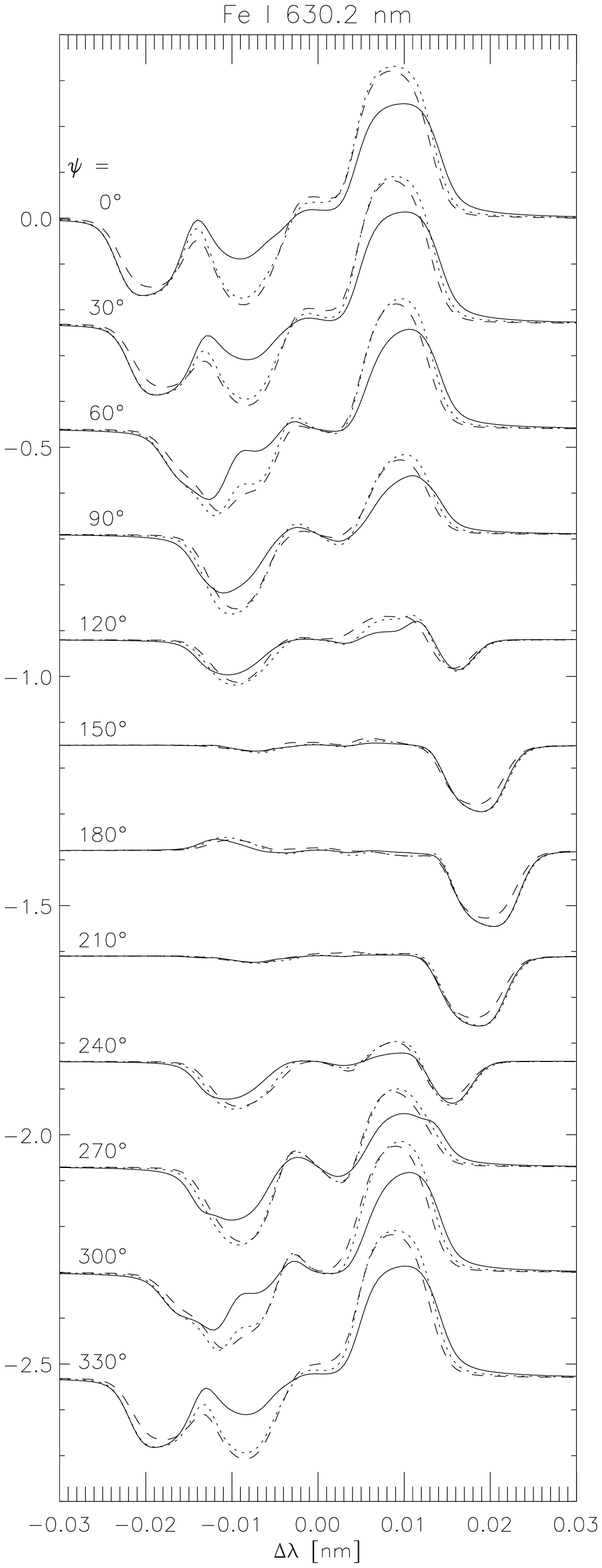}
}
\end{center}
\caption{\label{gfx_vstack_atm}
The effect of different background atmospheres. The left panel shows the Stokes-V profiles of the infrared line, the right panel the Stokes-V profiles of the visible line. The solid line corresponds to the JS94 model, the dotted line to the \cite{BellotRubio+al2006AA} model and the dashed line to the model of \cite{SchleicherPhD}.
}
\end{figure*}

\begin{figure}[ht]
\resizebox{\hsize}{!}{
\includegraphics{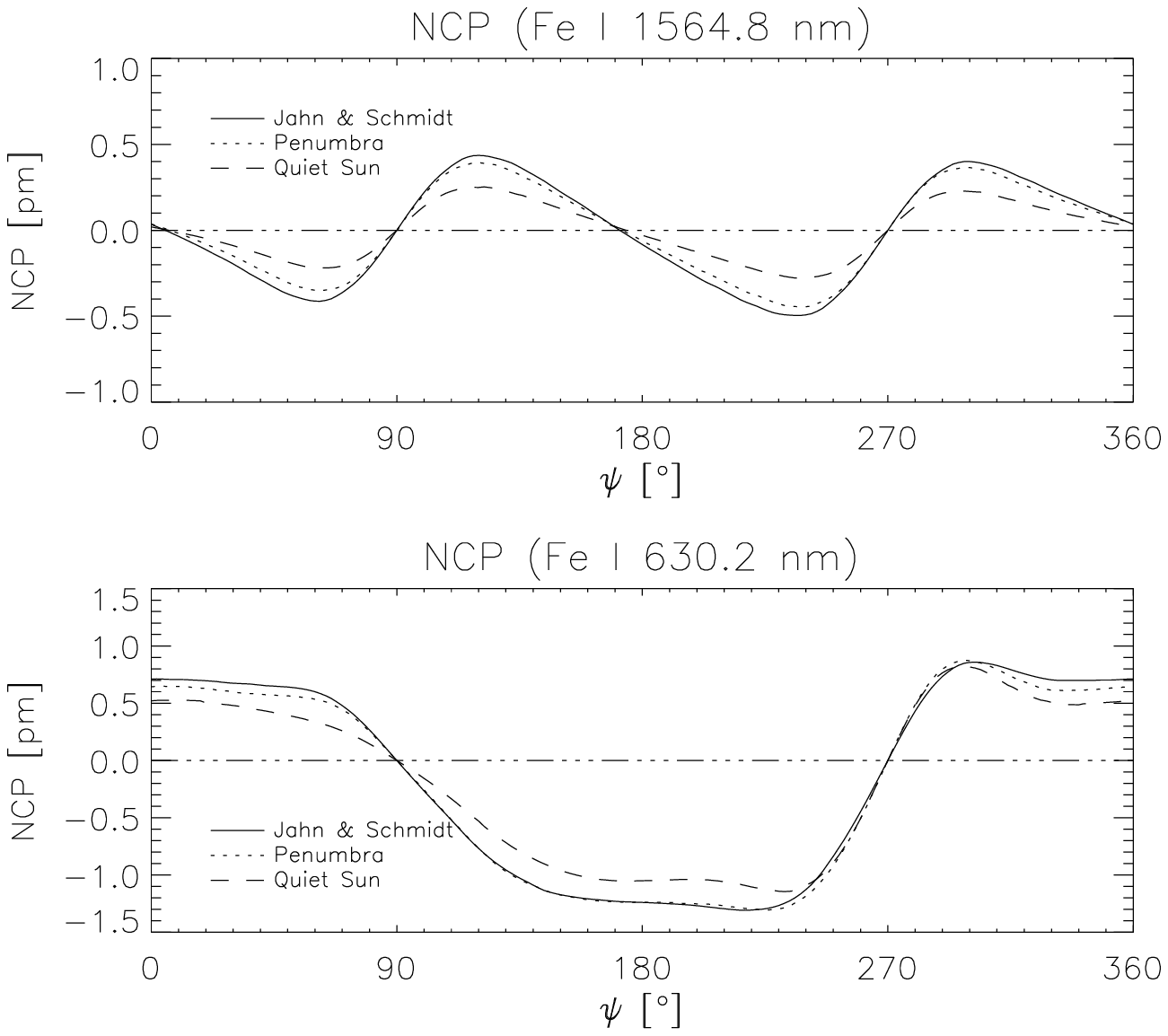}
}

\caption{\label{gfx_atm}
Azimuthal variation of the NCP for different background atmospheres.
The upper panel shows the Stokes-V profiles of the infrared line, the lower panel the Stokes-V profiles of the visible line. The solid line corresponds to the JS94 model, the dotted line to the one of \cite{BellotRubio+al2006AA} and the dashed line to the model of \cite{SchleicherPhD}.
}
\end{figure}

In order to check whether the atmospheric background model has a strong influence on the NCP, we calculated the same model with different background atmospheres. The remaining parameters ($r = 12\,000$\,km, $\alpha = 0^\circ$, $\theta = 30^\circ$, $\Delta B = 0$\,G) were kept constant. Figure~\ref{gfx_vstack_atm} shows stack plots of the resulting Stokes-V profiles and Fig.~\ref{gfx_atm} displays the corresponding NCP curves. 

In the case of the infrared line \ion{Fe}{1}~1564.8\,nm (left panel), the profiles look similar in shape for all three atmospheric models. The line depths are largest for the JS94 model, followed by the \cite{BellotRubio+al2006AA} model and the \cite{SchleicherPhD} model. This is simply a result of the temperature difference in the line-forming region between the three models: The JS94 model is the coolest model and thus absorbs more strongly than the warmer model of \cite{BellotRubio+al2006AA} and the even hotter model of \cite{SchleicherPhD}. This is consequently reflected in the NCP curves which are similar in shape for all three atmospheres, but slightly larger in amplitude the cooler the model is.

For the visible line \ion{Fe}{1}~630.25\,nm, the situation is slightly more involved. The right panel of Fig.~\ref{gfx_vstack_atm} shows significant differences in the shapes of the profiles on the center-side. To be more precise, the unshifted components of the Stokes-V profiles for the JS94 model differ from those of the other two models in the sense that both wings are less pronounced than for the other two atmospheric models.
The reason for this is that the JS94 model is a relatively simple theoretical one in which the temperature gradient almost vanishes above $\log(\tau) = -1$, i.e.\ directly above the flux tube. For this reason, the line formation nearly ceases above the flux tube which results in the odd shapes of the unshifted Stokes-V components. A closer look at the line formation process reveals that the Stokes-V profiles for $\psi = 0^\circ$ reach their final shape already around $\log(\tau) = -1.5$ for this model, while this state is only reached around  $\log(\tau) = -2.5$  and $\log(\tau) = -3$ for the models of \cite{BellotRubio+al2006AA} and \cite{SchleicherPhD}, respectively.

For the infrared line \ion{Fe}{1}~1564.8\,nm, the vanishing temperature gradient of the JS94 model above the flux tube does not significantly affect the shape of the Stokes profiles since this line is formed in deeper layers of the atmosphere.

 Despite the fact that the Stokes-V profiles vary for the different atmospheric models, the resulting NCP curves for the \ion{Fe}{1}~630.25\,nm line do not differ strongly. The slightly higher NCP for this model on the center-side penumbra can be attributed to the difference in line formation described above.
From these findings, we conclude that the choice of the atmospheric background model can significantly influence the shape of the Stokes profiles, but does not change the global characteristics of the resulting NCP curves for the tested models.

\begin{figure}[ht]
\resizebox{\hsize}{!}{
\includegraphics{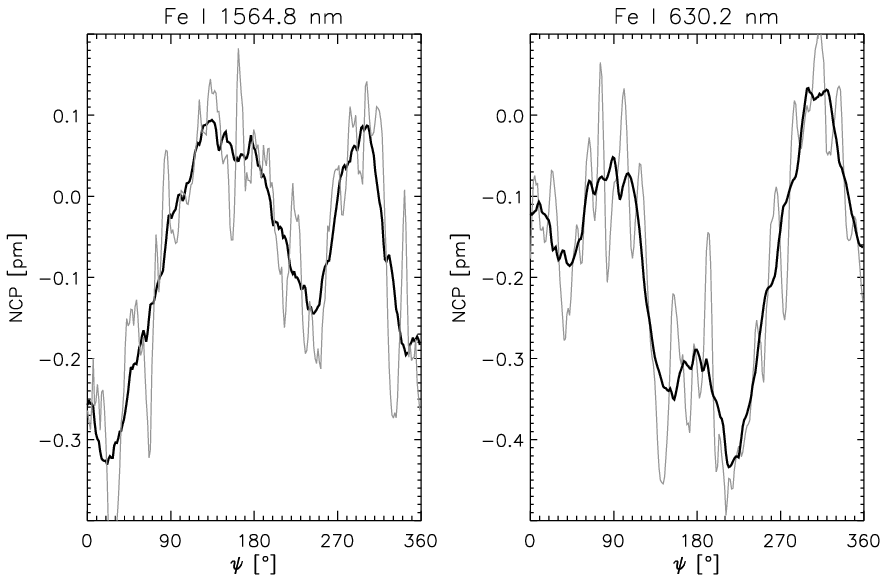}
}

\caption{\label{gfx_ncp_data_slice}
Observed azimuthal variation of the NCP for the infrared line (left panel) and the visible line (right panel). The thin lines show the raw data, the thick lines are smoothed with a boxcar average of $35^\circ$.}
\end{figure}

\noindent
For comparison, Fig.~\ref{gfx_ncp_data_slice} shows the observed azimuthal variation of the NCP for the infrared line \ion{Fe}{1}~1564.8\,nm and the visible line \ion{Fe}{1}~630.25\,nm. As noted above, the absolute magnitude of the observed NCP is lower than the one calculated from the synthetic profiles since the latter assume a filling factor of unity, i.e.\ each line of sight is passing through the center of a magnetic flux tube. The exact calibration of the zero-level of the observed NCP is very difficult, therefore Fig.~\ref{gfx_ncp_data_slice} shows the unshifted raw data.

\section{Summary}
We have presented a generalized geometrical model that embeds an arbitrarily shaped flux tube in a stratified magnetized atmosphere. The new model is a versatile tool to calculate the spectral signature of flux tubes in the penumbra and especially to make predictions about the flow speed and tube inclination from observed maps of the net circular polarization. From the first applications of the new model, we find that
\begin{itemize}
\item[(a)] The inclination of downflows in the outer penumbra must be shallower than approx. 15$^\circ$.
\item[(b)] Observing the limb-side NCP of sunspots in the \ion{Fe}{1}~1564.8\,nm line offers a promising way to identify a reduced magnetic field strength in flow channels.
\item[(c)] The choice of the background atmosphere can significantly influence the shape of the Stokes profiles, but does not change the global characteristics of the resulting NCP curves for the tested atmospheric models.
\end{itemize}

\noindent
In future work, we will compare synthetic NCP maps obtained from this forward-modeling approach with results from Stokes inversion maps of observational data in more detail. This will provide further insight into the structure of penumbral filaments and guide improvements in both modeling and inversion techniques.

\begin{acknowledgements}
The authors would like to thank W.~Schmidt and the anonymous referee for useful and constructive comments which helped to improve the quality of this paper.
C.~B. acknowledges a grant by the \emph{Deut\-sche For\-schungs\-ge\-mein\-schaft, DFG\/}.
\end{acknowledgements}

\bibliographystyle{aa}
\bibliography{aamnem99,polarization}

\begin{appendix}
\section{Model description}

\subsection{Defining the flux tube}
In order to construct a computationally efficient algorithm that is
able to calculate the physical properties along the
line-of-sight, we use the parametrized 3D model described below. In
contrast to an explicit 3D model where one generates a three-dimensional
box for all the physical properties and then intersects this box with a line-of-sight, our model generates the line-of-sight variables directly from
geometrical considerations.

\begin{figure}[ht]
\begin{center}
\includegraphics[width=0.8\columnwidth]{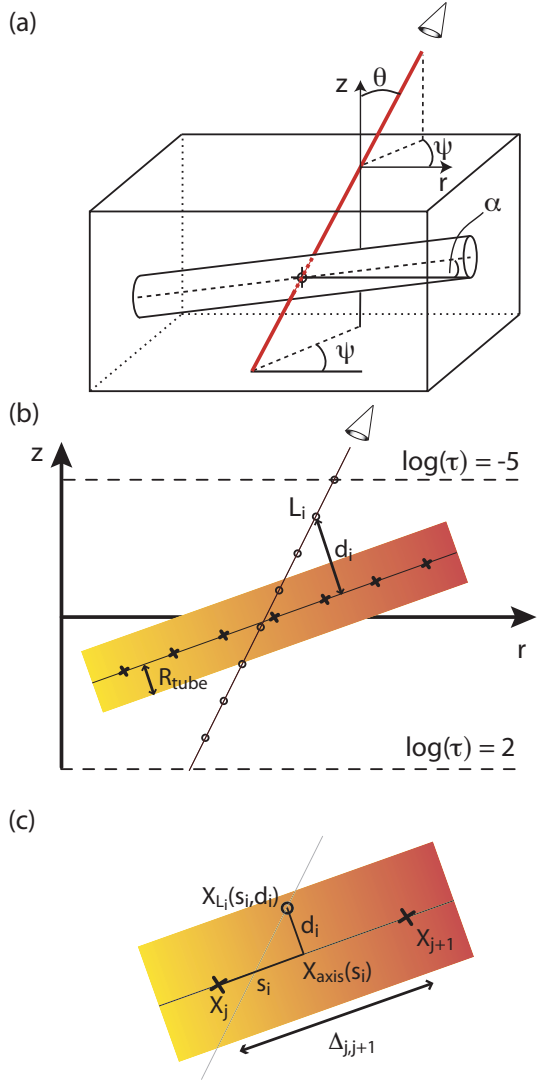}
\end{center}
\caption{\label{FIG1}Panel (a) shows a flux tube, inclined by an angle $\alpha$ and observed at a heliocentric angle $\theta$ and a spot angle $\psi$.
Panel (b) shows the flux tube intersected by the LOS, projected into the $(r,z)$-plane. For each point $L_i$ on the LOS, the minimal distance $d_i$ to the flux tube axis is calculated. If this distance is smaller than the local tube radius, the point $L_i $ lies inside the flux tube. Typically, the radiative transfer equation is integrated from an optical depth of $\log(\tau)=2$ out to  $\log(\tau)=-5$ (indicated as dashed lines for a vertical LOS). In panel (c), the relevant dimensions for the interpolation of an arbitrary physical quantity $X$ are depicted. First, a linear interpolation along the flux tube axis is carried out, thus leading to a value $X_{axis}(s_i)$ at distance $s_i$ from $X_j$. Second, starting from the value $X_{axis}(s_i)$ at the flux tube axis, one can prescribe an arbitrarily shaped radial dependence, which results in $X_{L_i}(s_i,d_i)$.}
\end{figure}

We start with a two-dimensional definition of the flux tube
axis, which is given by a polygon or spline function and is embedded
in the $(r,z)$-plane as shown in Figs.~\ref{FIG1}a and b. This figure shows an example of a straight tube with a constant inclination angle, $\alpha$, but in principle, $\alpha$ is variable along the tube. A complete set of physical quantities relevant for the radiative transfer code is assigned to each point along the tube axis. The local radius of the tube, $R_{tube}$, is calculated from the condition that the magnetic
flux, ${\cal F}$,  is constant along the tube,

\be
R_{tube}=\sqrt{\frac{{\cal F}}{\pi B}}\,.
\ee

\noindent
We then define the discretized LOS which intersects the tube at any desired point. The inclination $\theta$ of the LOS is given by the heliocentric angle and its azimuth $\psi$ corresponds to the spot angle. For all points $L_i$
($i=1,\ldots,N_{tube}$) along the LOS we calculate the nearest 3D
distance, $d_{L_i}$, to the tube axis as shown in Fig.~\ref{FIG1}b. If
$d_{L_i}<R_{tube}$, the point $L_i$ lies inside the flux tube, otherwise it is associated with the penumbral background atmosphere. Based on this distance-dependent binary decision, we assign the physical properties to each
point $L_i$ along the LOS.

\subsection{Assignment of the background properties}
\begin{figure}[ht]
\includegraphics[width=\columnwidth]{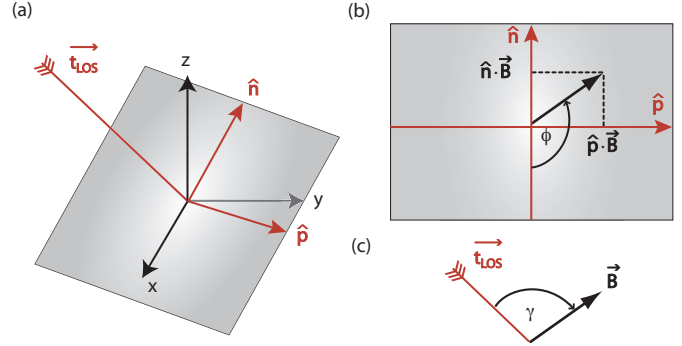}
\caption{\label{FIG2}(a) Illustration of the local coordinate system ($\hat{n},\hat{p}$), spanning the plane perpendicular to the line-of-sight. The azimuthal angle $\phi$ with respect to the line-of-sight is shown in (b), whereas (c) shows the definition of the inclination between the line-of-sight and the magnetic field.}
\end{figure}

Based on the coordinates $(r_i,z_i)$ of the points $L_i$ we can easily
interpolate the scalar quantities from tabulated background model
values (temperature $T$, pressure $p$, density $\rho$, magnetic field
strength $B$). In addition, one needs the
magnetic field azimuth $\phi$ and its inclination $\gamma$ with respect
to the LOS (see Fig.~\ref{FIG2}b and c). Let $\vec{t}_{LOS}$ be the
tangent vector along the LOS (directed into the spot) and $\vec{B}$
the magnetic field vector. The inclination $\gamma$ is then given by
(compare Fig.~\ref{FIG2}c)
\be \gamma = \arccos{\frac{\vec{B} \cdot
    (-\vec{t}_{LOS})}{|\vec{B} | |\vec{t}_{LOS} |}}.  
\ee 
The minus sign arises due to the fact that we set $\gamma=0^\circ$ when $\vec{B}$ and $\vec{t}_{LOS}$ are antiparallel, i.e.\ an inclination angle $\gamma \in [0^\circ,90^\circ)$ indicates that the magnetic field points into the hemisphere of the observer. We assume positive magnetic polarity, i.e.\ the magnetic field vector is parallel to the flow vector inside the flux tube and points out of the solar surface for the magnetic background field.

\noindent
In order to calculate the magnetic field azimuth $\phi$, we have to project the
magnetic field vector onto the plane perpendicular to the LOS as illustrated in Fig.~\ref{FIG2}a. We define $\hat{p}$ as the unit vector perpendicular
to $\vec{t}_{LOS}$ and the $z$-axis ($\hat{p}$ is consequently located
in the $(x,y)$-plane) and $\hat{n} $ as the unit vector normal on
$\vec{t}_{LOS}$ and $\hat{p}$. A closer look at Fig.~\ref{FIG2}b shows that $\phi$ is given by 
\be
\label{eq_phi}
\phi = 
\left\{ 
  \begin{array}{llll}
    &\arctan{(a/b)}+\pi/2 &  & a > 0 \wedge b > 0  \\
    -& \arctan{(a/b)} & \mathrm{for}\;& a > 0 \wedge b < 0 \\
    &\arctan{(a/b)}+3\pi/2 & & a < 0 \wedge b < 0 \\
    -& \arctan{(a/b)}+\pi & & a < 0 \wedge b > 0 \\

 \end{array} 
\right .\,,  \ee

\noindent
where $a = \vec{B} \cdot \hat{p}$ and $b = \vec{B} \cdot \hat{n}$.
\subsection{Assignment of the flux tube properties}
If a point $L_i$ lies inside the flux tube, the physical
properties are interpolated from the values provided along the flux
tube polygon. In particular, one first interpolates the quantity of
interest (here symbolically termed $X$) linearly along the flux tube
axis. If the two nearest flux tube neighbors of $L_i$ are the points
$j$ and $j+1$ with quantities $X_j$ and $X_{j+1}$, respectively, we obtain
for $X_{axis}$ in a distance $s_i$ from $j$ (compare Fig.~\ref{FIG1}b)
\be
X_{axis}(s_i) = X_j + \frac{s_i}{\Delta_{j,j+1}}(X_{j+1}-X_j) \,.
\ee
Second, we have the freedom to analytically prescribe an arbitrarily
shaped radial dependence {\it f} for $X$,
\be
X_{L_i}(s_i,d_i)=
f(X_{axis}(s_i),d_i) \,, 
\ee 
with the limitation that $f(X_{axis}(s_i),0)=X_{axis}(s_i)$. Such a dependence of the physical properties on the radius can e.g.\ be prescribed to model the cooling of a flux tube. In this study we do not investigate the impact of such radial dependences and use instead (as mentioned above) the thin flux tube approximation, where the physical variables are radius-independent,
\be
f(X_{axis}(s_i),d_i)= X_{axis}(s_i)\,.
\ee

\noindent
The vector-valued quantities (i.e.\ the velocity and the magnetic
field) are assumed to be oriented parallel to the flux tube axis and
their strength is interpolated from the values given on the axis in
the aforementioned manner. The projection onto the plane
perpendicular to the LOS is carried out analogously to the previously
described way (compare Fig.~\ref{FIG2} and Eq.~\ref{eq_phi}).

After all physical quantities have been projected onto the line-of-sight, they are passed to the radiative transfer code. Scanning the spot angle and the radial position of the line-of-sight within the penumbra, one can now easily obtain maps all four Stokes parameters or any derived quantities.

\end{appendix}

\end{document}